\documentclass{ws-ijmpcs}
\begin{document}

\markboth{Paola Grandi}
{GAMMA-RAYS FROM RADIO GALAXIES: FERMI-LAT OBSERVATIONS}

%
\catchline{}{}{}{}{}
%

\title{GAMMA-RAYS FROM RADIO GALAXIES: FERMI-LAT OBSERVATIONS}

\author{PAOLA GRANDI\\ ON BEHALF OF THE FERMI-LAT COLLABORATION}

\address{Istituto Nazionale di Astrofisica -IASFBO\\
Via Gobetti 101, I-40129 Bologna, Italy\\
grandi@iasfbo.inaf.it}

\maketitle
\begin{history}
\end{history}

\begin{abstract}
We review the high energy properties of Misaligned AGNs associated
with $\gamma$-ray sources detected by {\it Fermi} in 24 months of
survey.  Most of them  are nearby  emission low power radio galaxies
(i.e FRIs) which probably have structured jets. On the contrary, high
power radio sources (i.e FRIIs)  with GeV emission  are rare. The
small number of FRIIs does not seem to be related to their higher
redshifts.  Assuming proportionality between the radio core flux and the $\gamma$-ray flux, several of them  are expected to  be bright enough to be detected above 100 MeV in spite of their distance.
We suggest that beaming/jet  structural  differences  are responsible for  the detection rate discrepancy observed between FRIs and FRIIs.

\keywords{gamma rays: observations - galaxies: active - galaxies: jets }
\end{abstract}

\ccode{PACS numbers: 95.85.Bh, 95.85.Pw; 98.54.Gr}

\section{Introduction}	
\begin{figure}[h]
\centerline{\psfig{file=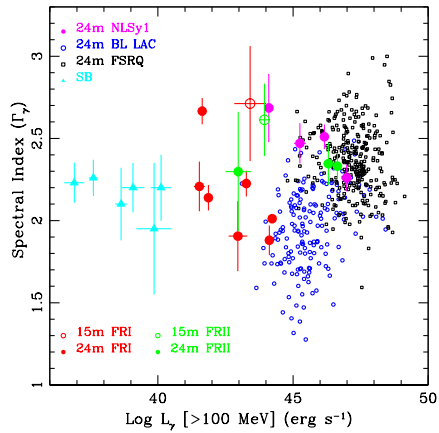,width=6.0cm}}
\vspace*{6pt}
\caption{Spectral slope versus $\gamma$-ray luminosity of the sources detected by the LAT in 24 months. The small group of SBs, NLSy1s and MAGNs  preferentially fill the plot region characterized by lower luminosities.\label{F1} }
\end{figure}

The Large Area Telescope (LAT\cite{Atwood2009}) on-board the $\gamma$-ray  ${\it Fermi}$ satellite
 detected more than one thousand extragalactic
 sources\cite{Ackerman2011} in two years of survey.  The majority are
 blazars [i.e. BL Lacs and Flat Spectrum Radio Quasars (FSRQs)], confirming that AGNs with the jet oriented in the direction of the observer are the brightest GeV sources. Only a handful of LAT objects have  different counterparts, i.e. Starburst galaxies (SBs), Narrow Line Seyfert 1 galaxies (NLSy1s) and Misaligned AGNs (MAGNs)\footnote{MAGNs are Radio Galaxies and Steep Spectrum Radio Quasars, i.e. Radio Loud objects with steep radio spectra ($\alpha>0.5$) and/or showing possibly symmetrical extension in radio maps.}.  The state of the art of extragalactic observations after 24 months of {\it Fermi} activity is shown in 
Fig.~\ref{F1} where the $\gamma$ spectral slope ($\Gamma_ {\gamma}$) of blazars and non-blazar objects is plotted as a function of  the luminosity ($L_{>100~MeV}$).

In spite of their small number, the non-blazar $\gamma$-ray emitters are extremely appealing, as they  offer a powerful physical tool in approaching  the high energy phenomena.  Starburst Galaxies, for example, are a  laboratory for investigating cosmic ray acceleration\cite{sb1}.
The high energy jet emission discovered in NLSY1s is questioning the paradigm according to which radio-loud AGNs are only hosted in elliptical galaxies\cite{Fos011}. The detection of $\gamma$-ray photons in MAGNs is invaluable in revealing  the jet structure complexity\cite{Abdoa2010}. 

\section{$\gamma$-ray properties of Misaligned AGNs}

It is widely accepted that Doppler boosting effects favor the observations of blazars at very high energies.
The jet flux amplification (A=$F_{\rm observed}/F_{\rm intrinsic}$)  is proportional to a power of the kinematic Doppler factor $\delta$ defined as $\delta=[\Gamma(1-\beta cos\theta)]^{-1}$, being $\beta$ the bulk velocity in units of the speed of light, $\theta$ the angle between the jet and the line of sight and $\Gamma=(1-\beta^2)^{-1/2}$ the Lorentz factor. 
If the observed  flux is produced through synchrotron emission, $A=\delta^{3+\alpha}$, where $\alpha$ is the synchrotron power law index. 
The flux enhancement is strongly dependent on the viewing angle and decreases very rapidly for $\theta > 8^{\circ}$-$10^{\circ}$.
For this reason, large inclination Radio Loud AGNs were not considered
appealing  $\gamma-$ray targets before the {\it Fermi} launch.
Only some Fanaroff-Riley I and  a few Fanaroff-Riley II  radio
galaxies (RGs)\footnote{Following the Fanaroff-Riley\cite{FR} classification, radio galaxies  belong to  
two radio morphological classes corresponding to  edge-darkened (FRI) and edge-brightened (FRII) objects, with FRII objects 
being more powerful  (P$_{178~MHz}>10^{25}$ W Hz$^{-1}$ sr$^{-1}$) than FRI radio galaxies. 
FRIs are considered the parent population of  BL Lacs, while FRIIs are
thought to be Flat Spectrum Radio Quasars (FSRQs) with jet axis not aligned with the line of sight.}
were suggested to have fluxes above the LAT sensitivity threshold\cite{ghi05}$^{-}$\cite{gp07}.

The LAT detection of eleven  objects\cite{Abdoa2010} in only 15 months of GeV sky exploration  has successively confirmed MAGNs to be a new class of $\gamma$-ray emitters.  
Among MAGNs, only three source, i.e. 3C111, Centaurus A and NGC6251, have been proposed as candidates by the previous $\gamma$-ray telescope EGRET.  The other ones represent a new discovery.   Most of the sources of the 15month-MAGN sample are faint  ($F(>0.1~GeV) \sim10^{-8}$  photons $cm^{-2}$ $s^{-1}$) and have steep power law spectra ($\Gamma>2.4$).
This is  in general agreement with the AGN Unified Models that assume MAGNs to be a de-boosted version of blazars. 
Because of their faintness, variability studies are not conclusive. Only in one case, NGC1275, flux and spectral changes could 
be statistically ascertained on time scale of months\cite{ka10}. As a consequence, establishing where  the  $\gamma$-rays are produced  is a difficult task. The variability of NGC1275 seems to suggest a sub-pc scale ($<10^{18}$ cm) emission region , but the discovery of $\gamma$-ray emission from the radio lobes of Centaurus A\cite{Abdo10b} shows that extranuclear extended  kpc regions can also be  sources of high energy photons.

\section{FRII sources: an elusive $\gamma$-ray population}

\begin{figure}[h]
\centerline{\psfig{file=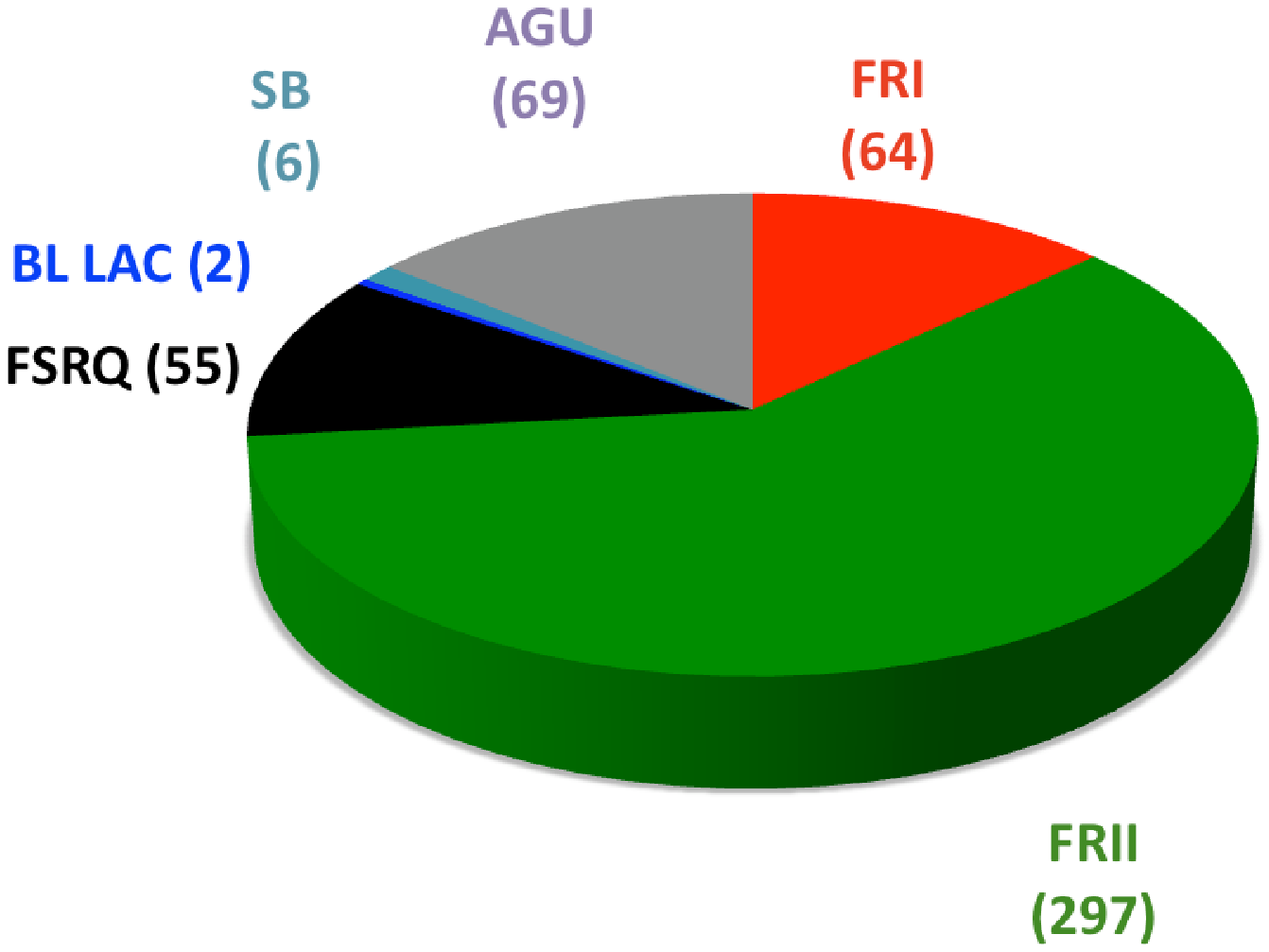,width=4.7cm} 
\psfig{file=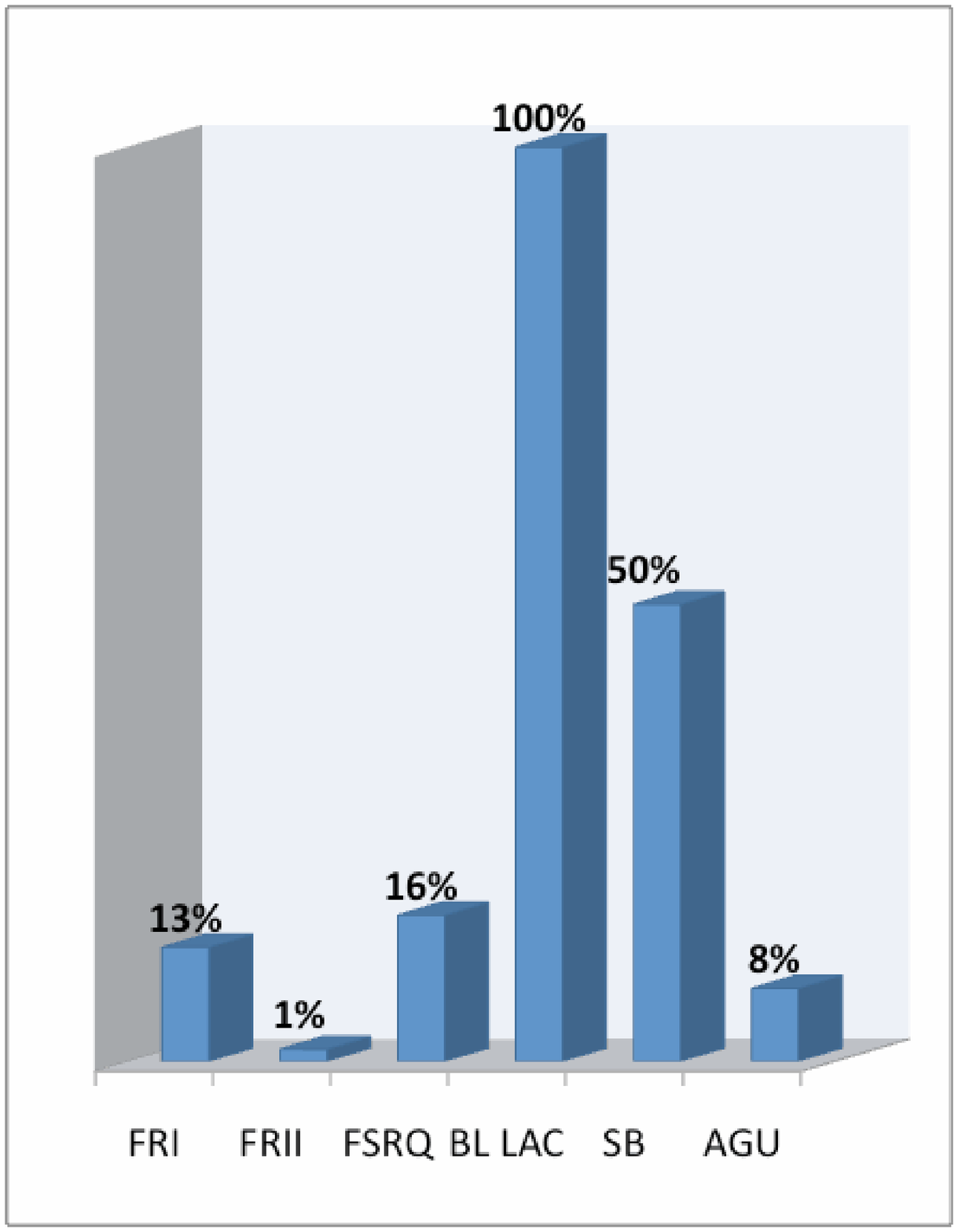,width=4.7cm}}
\vspace*{6pt}
\caption{{\it Left Panel} - Number of objects belonging to each class of radio sources. \label{F2a}.
{\it Right Panel} - $\gamma$-ray detection rate of each class of radio sources\label{F2b}. The percentages of FRIs and FRIIs in the radio source sample are 13$\%$ and 60$\%$, respectively, compared to the detection rate of 1$\%$ and 13$\%$, respectively.}
\end{figure}

The 15month-MAGN sample is dominated by nearby FRI radio galaxies. FRII radio sources appear to be more elusive objects, as also attested by a successive 18 month-study searching for gamma counterparts of Broad Line Radio Galaxies (mostly showing a FRII radio morphology)\cite{kata11}. 
In order to investigate this aspect, we take advantage of the publication of the second AGN LAT catalogue (2LAC\cite{Ackerman2011}) to enlarge the sample of MAGNs.
At first the 3CR\cite{sp85}, the revised 3CRR catalog\cite{la83}, the Molonglo Southern 4 Jy Sample (MS4)\cite{bh06a}$^{-}$\cite{bh06b} and the 2 Jy sample\cite{mo93} were combined (only one entry was considered for the sources listed in more catalogs) to have a large number of radio sources with optical and radio classifications. 
The demography of the combined sample (3C-MS4-2Jy sample)  is
represented in Fig.~\ref{F2a} - {\it left panel}. Not surprisingly,
the FRII is the most crowded class. 
We intentionally considered the low radio frequency 3CR/3CRR and MS4
samples because they preferentially select  radio sources characterized by
steep-spectrum synchrotron emission from extended lobes\footnote{Unlike the 3CR, 3CRR and MS4 catalogs, the 2Jy sample
  is a 5GHz survey able to easily detect bazars. It is useful in comparing MAGN and blazar properties}.  Then this big radio sample was cross-correlated with the 2LAC catalog.
The result is shown in  Figure~\ref{F2b} -{\it right panel}, where the percentages of FRIs, FRIIs, FSRQs,  BL Lacs, SBs and AGUs (AGNs with unknown classification) with a $\gamma$-ray association are reported. Some FRI and FRII sources, that are not in the second year catalog (because below the 2LAC adopted TS$>25$ threshold) but were detected on shorter integration (12-18 months) time,  are considered in calculating the fraction of detections.
BL Lacs and Starbursts have the highest probability to be detected, while FRIs and FSRQs have similar  detection rates. {\it Fermi} appears to be  almost blind to misaligned powerful radio  sources. 
Only $1\%$ of FRIIs is  visible at GeV energies, despite they are  the most numerous objects in the 3C-MS4-2Jys combined catalog.\\

\subsection{Possible interpretations}
\begin{figure}[h]
\centerline{\psfig{file=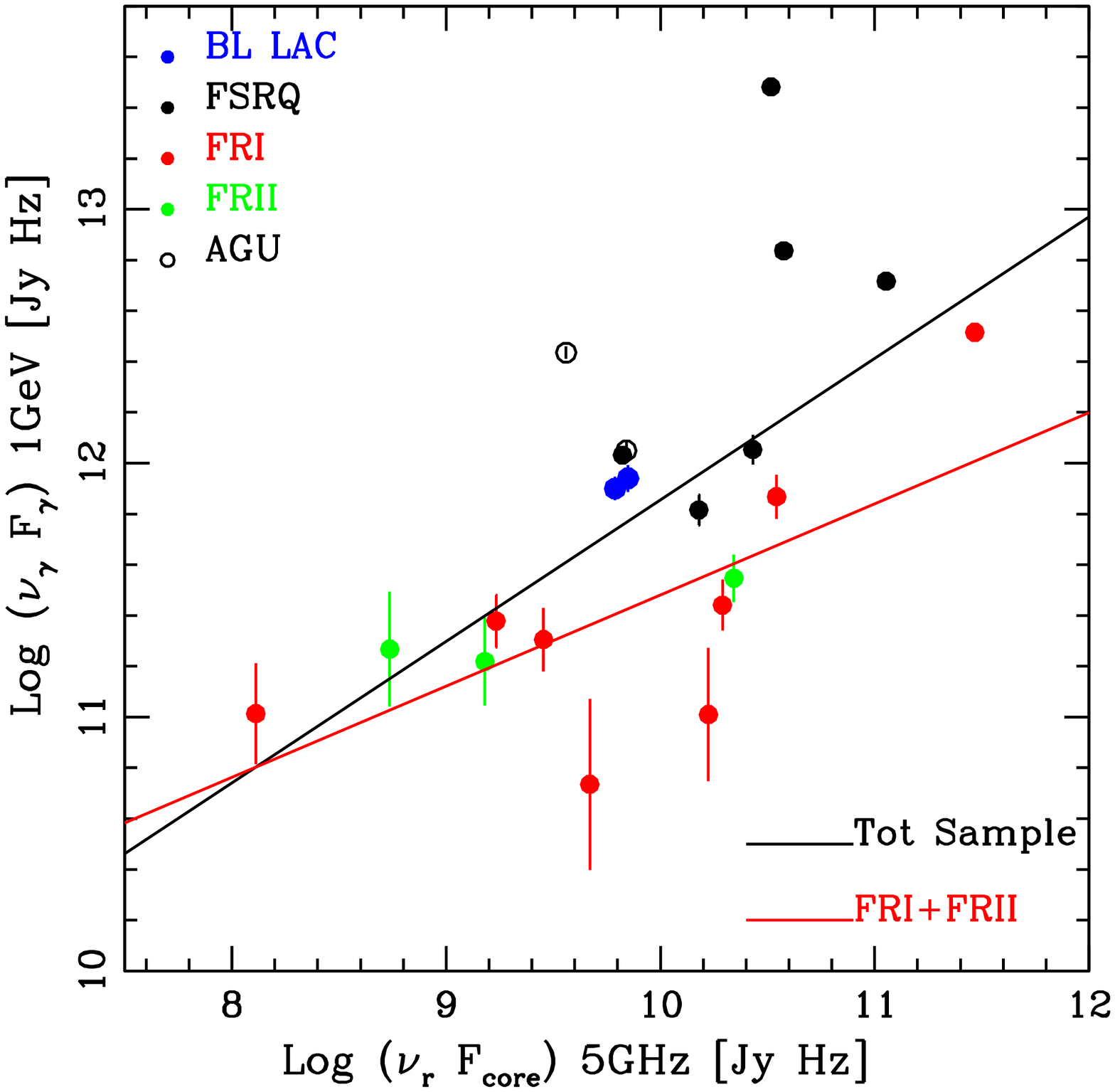,width=5.0cm} 
\psfig{file=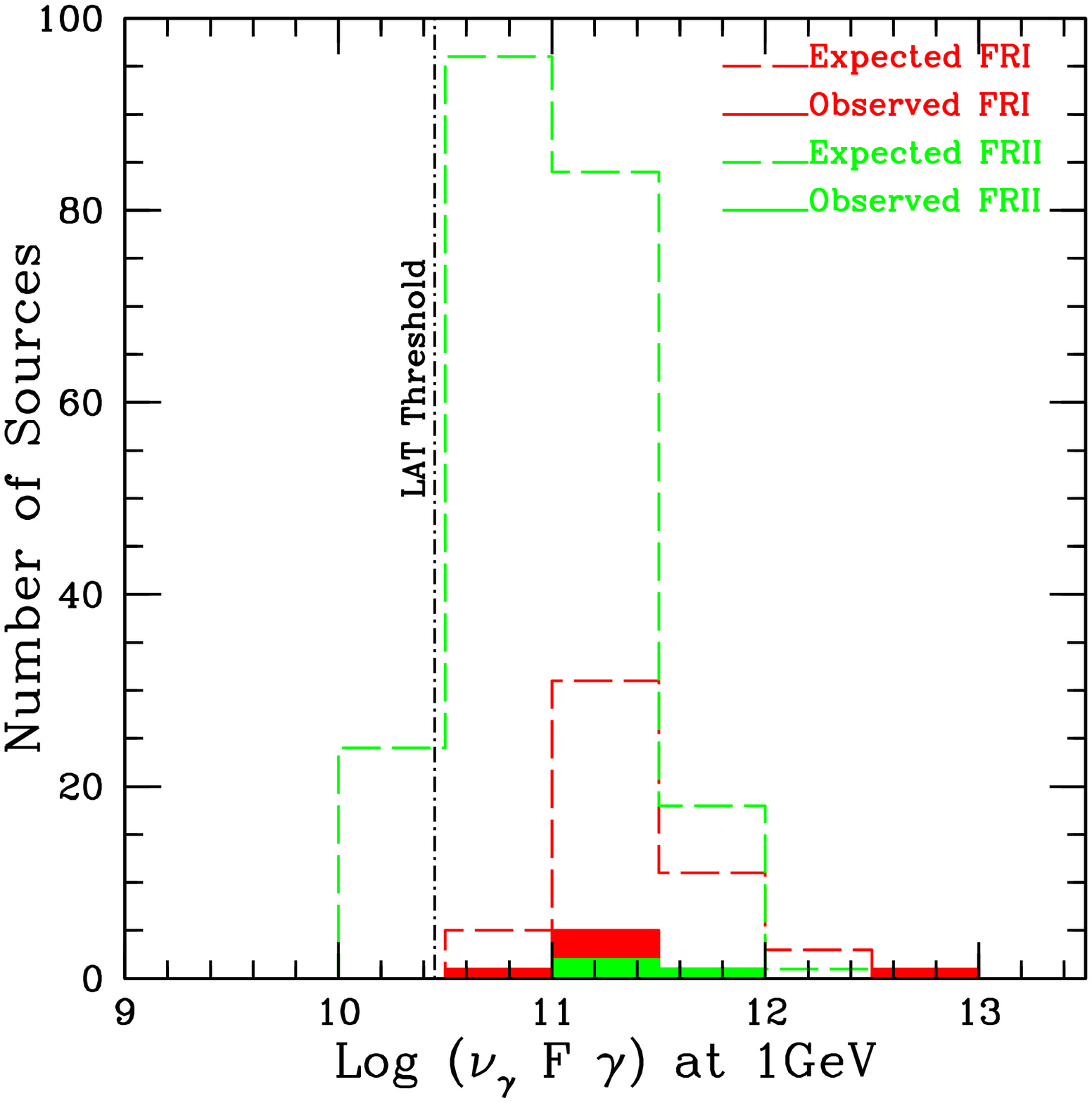,width=5.0cm}
}
\vspace*{6pt}
\caption{{\it Left Panel} - Radio-Gamma correlation of the 3C-MS4-2Jy
  objects  with a $\gamma$-ray  association (black line). A positive
  (flatter) trend is also present when only FRI and FRI sources are
  taken into account (red line)\label{F3}. 
{\it Right Panel} -  Histogram of the predicted fluxes for all the FRI (red) and FRII (green) radio sources belonging to the 3C-MS4-2Jy sample (dashed lines). For comparison, the real observed MAGNs are also shown. Many FRIIs are expected to be bright enough to be detected by {\it Fermi}
\label{F4}.
} 
\end{figure}

The simplest interpretation accounting for the different FRI and FRII detection rates is that we are  losing radio powerful MAGNs  because more distant 
(and therefore fainter). In order to verify this possibility, we attempted to estimate the $\gamma$-ray fluxes of the MAGNs for which no GeV association was found. 
We evaluated  the 1 GeV  flux  of all the undetected sources,
using (in a first approximation) the positive correlation observed between the 5~GHz core fluxes and the 1 GeV flux of the 3C-MS4-2Jy sources  with  a $\gamma$-ray counterpart (see Fig~\ref{F3} {\it left panel}). We implicitly assumed that $\gamma$-ray emission occurs in similar physical conditions in all AGNs.
In agreement with the Spearman test, this correlation has less than a $1\%$ probability of being due to chance alone and  it is still present (chance probability less than $3\%$)  when only FRIs and FRIIs are taken into account. Incidentally, we note that a linear gamma-radio relation, considering larger samples of AGNs,   has also been reported  by other authors\cite{ghirlanda2011}$^{-}$\cite{Ackermann11}.  In the histogram of Fig.~\ref{F4}-{\it right panel} the predicted and observed gamma fluxes for both  FRI and FRII sources are compared. Here only the results  based on the MAGNs  (red line in Fig. \ref{F3}-{\it left panel}) correlations are presented.  However a similar histogram was obtained 
considering the correlation including  blazars and AGUs (black line in Fig. \ref{F3}-{\it left panel}). 
Distance/faintness effects can not explain the  FRII scarcity  in the $\gamma$-sky. As shown in Fig.~\ref{F4}-{\it right panel},
many FRIIs are expected to be as bright as FRIs and above the LAT sensitivity threshold. 
Different effects must then be invoked to justify the low probability of detecting  high powerful radio sources above 100 MeV. 

It is possible that one of our implicit assumptions, i.e. same flux boosting factor at low and high frequencies in both FRIs and FRIIs, is  too simple.
For example, the Doppler boosting is stronger and the beaming cone
narrower compared to synchrotron processes if the emission is due to
Compton scattering of external photons (EC) in the
jet\cite{dermer95}. If the high energy emission is dominated (as is
likely) by EC processes in powerful radio sources and by SSC processes
in low power radio galaxies, a beaming difference  could account for
the smaller number of FRIIs when compared to FRIs. 

It is also probable that  one-zone homogeneous SSC/EC models are an oversimplified interpretations of  the jet emission\cite{reynoso11}.
Spectral Energy Distributions (SEDs) of FRI radio galaxies such as NGC1275\cite{Abdo2009b}, M87\cite{Abdo2009c} and NGC6251\cite{migliori}
are consistent with an SSC model with Lorentz  factors ($\Gamma
\lesssim  3$),  much lower than typical values found in
models of BL Lac objects. This is  in apparent contrast with the AGN Unification Scheme\cite{migliori}, unless the assumption of a one-zone homogeneous SSC emitting region is relaxed
and a structured jet is assumed (as also attested by recent multifrequency observations and polarization studies\cite{gomez}).
Among the proposed scenarios, decelerating jet flow\cite{Georganopoulos2003} and spine-layer jet\cite{ghi05} models seem to be promising in describing the jet complexity\cite{migliori}. In these models, an efficient (radiative) feedback between different regions increases the IC emission.
In one case, the jet is assumed to be decelerating; in the other one, it is supposed to be fast in the inner part  and slow in the external envelope. 
The discrepancy between  BL Lac and FRI bulk velocities can then be reconciled. In large inclination RGs we could be observing  the external sheets, while in BL Lacs the jetted radiation is directly coming from the inner fast spine.
Within this context, the deficit of $\gamma$-ray  photons in FRIIs (i.e. in  AGNs having powerful accretion disks and poorer hot gas
environments\footnote{As well known, the low luminosity FRI sources are on average in richest clusters than high powerful FRII sources\cite{pp88} } )could be due to less prominent (or absent) external layers  and/or less efficient deceleration processes.
In order to check this possibility, accurate modeling of FRII Spectral Energy Distributions are under investigation.

\section{Summary and Conclusions}
Although MAGNs are a minority in the GeV sky,  their study is particularly fruitful in providing general insights into the jet structure. We find that more then $10\%$ of the FRI radio galaxies of the radio sample obtained combining the 3CR, 3CRR, MS4 and 2Jy catalogs have $\gamma$-ray associations. Their LAT detections seem to be  favored by the presence of different velocity zones in the jet. Only $1\%$ of FRIIs, belonging to the same 3C-MS4-2Jy sample, are visible above 100 MeV.  The small number of detections does not seem to be due to their larger distances. More likely, it depends on less favorable  jet properties.

\footnotesize
\section*{Acknowledgments}
This contribution has greatly benefited  from the sharing of ideas with C. Dermer, L. Maraschi and G. Ghisellini .
P.G. would like to thank G. Palumbo for critical reading of the manuscript,  E. Torresi, G. Migliori  for stimulating discussions. and V. Bianchin for help with IDL.

\end{document}